\documentclass[runningheads]{llncs}

\usepackage[T1]{fontenc}
\usepackage{graphicx} 

\usepackage{array}
\usepackage[nocompress]{cite}
\usepackage{subcaption}
\usepackage{doi}
\usepackage{tabularx}
\usepackage{float}
\usepackage{xspace}

\usepackage{color}

\usepackage[dvipsnames]{xcolor}

\usepackage{algorithm2e}
\usepackage{amssymb}
\usepackage{complexity}
\usepackage{physics}
\usepackage{siunitx}
\usepackage{stackengine}
\usepackage{tikz}
\usepackage{nicematrix}
\usetikzlibrary{calc, positioning, shapes}
\pgfdeclarelayer{nodelayer}
\pgfdeclarelayer{edgelayer}
\pgfsetlayers{nodelayer,edgelayer}

\newcommand{\bsset}[1]{\begin{pgfonlayer}{nodelayer}
        \node [] (5) at (-1.75, 1) {};
        \node [] (9) at (1.75, 1) {};
        \node [] (10) at (-1.75, -1) {};
        \node [] (11) at (1.75, -1) {};
        \node [] (12) at (-0.5, 0) {};
        \node [] (12a) at (-0.5, 0.2) {};
        \node [] (12b) at (-0.5, -0.2) {};
        \node [] (13) at (0.5, 0) {};
        \node [] (14) at (0, 0.625) {#1};
        \node [] (15) at (-2.25, 1) {};
        \node [] (16) at (-2.25, -1) {};
        \node [] (17) at (2.25, -1) {};
        \node [] (18) at (2.25, 1) {};
        \coordinate (.tl) at (5);
        \coordinate (.bl) at (10);
        \coordinate (.tr) at (9);
        \coordinate (.br) at (11);
    \end{pgfonlayer}
    \begin{pgfonlayer}{edgelayer}
        \draw [in=180, out=0] (5.center) to (13.center);
        \draw [in=0, out=-180] (13.center) to (10.center);
        \draw (12.center) to (13.center);
        \draw [in=-180, out=0] (12.center) to (9.center);
        \draw [in=180, out=0] (12.center) to (11.center);
        \draw (15.center) to (5.center);
        \draw (10.center) to (16.center);
        \draw (11.center) to (17.center);
        \draw (9.center) to (18.center);
        \draw[fill=black!10] (12a.center) rectangle (13.center);
        \draw[fill=black] (12b.center) rectangle (13.center);
    \end{pgfonlayer}}

\newcommand{\bsrset}[1]{\begin{pgfonlayer}{nodelayer}
        \node [] (5) at (-1.75, 1) {};
        \node [] (9) at (1.75, 1) {};
        \node [] (10) at (-1.75, -1) {};
        \node [] (11) at (1.75, -1) {};
        \node [] (12) at (-0.5, 0) {};
        \node [] (12a) at (-0.5, 0.2) {};
        \node [] (12b) at (-0.5, -0.2) {};
        \node [] (13) at (0.5, 0) {};
        \node [] (14) at (0, 0.625) {#1};
        \node [] (15) at (-2.25, 1) {};
        \node [] (16) at (-2.25, -1) {};
        \node [] (17) at (2.25, -1) {};
        \node [] (18) at (2.25, 1) {};
        \coordinate (.tl) at (5);
        \coordinate (.bl) at (10);
        \coordinate (.tr) at (9);
        \coordinate (.br) at (11);
    \end{pgfonlayer}
    \begin{pgfonlayer}{edgelayer}
        \draw [in=180, out=0] (5.center) to (13.center);
        \draw [in=0, out=-180] (13.center) to (10.center);
        \draw (12.center) to (13.center);
        \draw [in=-180, out=0] (12.center) to (9.center);
        \draw [in=180, out=0] (12.center) to (11.center);
        \draw (15.center) to (5.center);
        \draw (10.center) to (16.center);
        \draw (11.center) to (17.center);
        \draw (9.center) to (18.center);
        \draw[fill=black] (12a.center) rectangle (13.center);
        \draw[fill=black!10] (12b.center) rectangle (13.center);
    \end{pgfonlayer}
}

\tikzstyle{phase}=[draw,shape=rectangle,minimum height=.8cm,fill=black!10]
\tikzset{
    bs/.pic={\bsset{$\theta$}},
    bsr/.pic={\bsrset{$\theta$}},
    ps/.pic={
    \begin{pgfonlayer}{nodelayer}
        \node [] (5) at (-1.1, 0) {};
        \node [] (9) at (1.1, 0) {};
        \node [style=phase] (11) at (0, 0) {\small $\psi$};
    \end{pgfonlayer}
    \begin{pgfonlayer}{edgelayer}
        \draw (5.center) to (11);
        \draw (11) to (9.center);
    \end{pgfonlayer}
    }
}

\RestyleAlgo{ruled}
\newcolumntype{H}{>{\setbox0=\hbox\bgroup}c<{\egroup}@{}}
\newcommand{\nat}{\mathbb{N} \cup \{0\}}
\newcommand{\posrat}{\mathbb{Q}^{+} \cup \{0\}}
\newcommand{\iu}{\mathrm{i}}

\newcommand{\probamp}{\alpha}
\newcommand{\compzero}{\mathtt{0}}
\newcommand{\compone}{\mathtt{1}}
\newcommand{\fockset}{\mathcal{F}}
\newcommand{\fock}[1]{\ket{#1}_{\fockset}}
\newcommand{\rail}[2]{\fock{#1}\fock{#2}}

\newcommand{\numaux}{m_a}
\newcommand{\coinbasis}{\mathcal{C}}
\newcommand{\qubitbasis}{\coinbasis_{qc}}
\newcommand{\fockbasis}{\mathcal{B}}
\newcommand{\Ulin}{\hat{U}}
\newcommand{\Ufock}{U_{\fockset}}
\newcommand{\creation}[1]{\hat{a}_{#1}^{\dagger}}

\newcommand{\sat}{\texttt{sat}}
\newcommand{\unsat}{\texttt{unsat}}
\newcommand{\unkown}{\texttt{unkown}}
\newcommand{\txtdsat}{$\delta$-\sat{}}
\newcommand{\dsat}{\delta\text{-\sat{}}}
\newcommand{\doptimal}{$\delta$-\texttt{optimal}}
\newcommand{\unchecked}{\texttt{unchecked}}
\newcommand{\infeasible}{\texttt{infeasible}}
\newcommand{\approximate}{\texttt{approximate}}

\newcommand{\ub}[1]{#1_{ub}}
\newcommand{\lb}[1]{#1_{lb}}

\newcommand{\eg}{\emph{e.g.},\xspace}
\newcommand{\ie}{\emph{i.e.},\xspace}
\newcommand{\etc}{\emph{etc}.}

\newcommand{\frob}[1]{\norm{#1}_F}

\usepackage{totcount}
\newtotcounter{todonum}

\title{Finding Photonics Circuits via $\delta$-weakening SMT}

\author{Marco Lewis\orcidID{0000-0002-4893-7658} \and Benoît Valiron\orcidID{0000-0002-1008-5605}}
\institute{Université Paris-Saclay, CNRS, CentraleSupélec, ENS Paris-Saclay, Inria,\\ Laboratoire Méthodes Formelles, 91190, Gif-sur-Yvette, France}

\authorrunning{M. Lewis and B. Valiron}

\begin{document}

\maketitle


\begin{abstract}
    For quantum computers based on photonics, one main problem is the synthesis of a photonic circuit that emulates quantum computing gates.
    The problem requires using photonic components to build a circuit that act like a quantum computing gate with some probability of success.
    This involves not only finding a circuit that can correctly act like a quantum gate, but also optimizing the probability of success.
    Whilst many approaches have been given in the past and applied to specific gates, they often lack ease of reusability.
    We present a tool that uses dReal, a $\delta$-weakening SMT solver, to find such photonic circuits, optimize the likelihood of occurring, and provide some guarantee that the result is optimal.
    We demonstrate the usage of our tool by recreating known results in the literature, extending upon them, and presenting new results for Givens rotation gates.
\end{abstract}


\section{Introduction}
\label{sec:intro}
The field of quantum computing has been making strides in recent years due to developments of different quantum computers.
These devices are developed using a variety of different physical techniques, such as ion trap~\cite{Bernardini24}, photonic devices~\cite{Kok07}, topological qubits~\cite{Kitaev03,Hormozi07}, \etc{}
Many of the techniques directly implement quantum computing gates from theory onto the physical devices.
However, some techniques require quantum gates to be implemented using basic building blocks.
Thus, the problem arises of how to synthesize a circuit of the basic building blocks that implements the desired quantum gate.

This is the case for quantum computers implemented using photonics, or linear optics.
The usage of linear optics to perform quantum computation was first introduced by Knill et al.\cite{Knill2001}.
Linear optics consists of wires that photons can be sent down and uses a variety of components (beam splitters, phase shifters, emitters, detectors, \etc{}) to implement unitary quantum gates.
The main difficulties of representing gates on a linear optical device is that there are many ways to represent the gate, and each representation has a probability to act as the desired gate and a probability that it acts in some other way.
Thus, whilst it is important to find a representation, it is also important to find the representation that is optimal, \ie{} has the highest probability of performing the desired operation.


In the past, various techniques have been used to find linear optical circuits to represent a quantum gate.
These include using computational tools to solve and simplify expressions~\cite{Knill02}, building a circuit~\cite{Ralph02,Liu23}, and through different decomposition schemes~\cite{Reck94,Clements16,Li2022,Alessio24}.
With these past techniques, it is usually the case that they return a circuit that implements the desired gate, but it does not guarantee optimality.

The usage of SMT and SAT solvers to perform forms of synthesis or optimization for quantum computing has been investigated previously.
Many approaches investigate the reduction in the depth of a quantum circuit~\cite{}, reducing the number of certain types of gates in a circuit (such as $CNOT$ gates)~\cite{}, and the depth of certain gates~\cite{Jakobsen25}.
Our usage of SMT solvers to investigate the synthesis of photonics circuits is a new line of research that has not been investigated before.

In this work, we present a search technique based on SMT solvers~\cite{SMTHandbook} to synthesize correct and optimal circuits for quantum gates.
We introduce the synthesis problem for linear optics and present the theory behind the developed search technique.
Our search method is primarily based on using a $\delta$-weakening SMT solver~\cite{deltasat,dreal} to find an approximation of a circuit (although the search can be adapted to standard SMT solvers).
This approximation is refined to an exact representation that can be implemented on a linear optics device.
These techniques have been implemented\footnote{Available at \url{https://doi.org/10.5281/zenodo.17116446}}, and we are able to check against known results in the literature and find new results.
Notably, our technique is automated in that one can input a quantum computing gate and the setup of the linear optics circuit, and the search is automatically performed.
This makes the technique generalisable; other techniques often require some work to apply it to different setups.


\section{Background}
\label{sec:background}
We provide a comprehensive background to quantum computing and linear optics in Sections~\ref{bg:quantumcomp} and \ref{bg:optics} respectively.
A full introduction to linear optics and its application to quantum computing can be found in~\cite{Kok07}.
For a summary of the problem we are trying to solve, see Section~\ref{bg:problem}.

\subsection{Quantum Computing}
\label{bg:quantumcomp}
In quantum computing, states are normally described by a complex vector and the base unit of information is the qubit.
A qubit resides in the unit circle of $\mathbb{C}^2$ and consists of the computational basis states $\ket{\compzero} = (1, 0)^\top$ and $\ket{\compone} = (0,1)^\top$.
Thus, a valid quantum state for a qubit can be written as $\ket{\phi} = \alpha_0\ket{\compzero} + \alpha_1\ket{\compone}$, where $\alpha_i \in \mathbb{C}$ and $\abs{\alpha_0}^2 + \abs{\alpha_1}^2 = 1$.

Qubits can be combined together using the tensor product.
Thus, an $n$-qubit system resides in the unit circle of $\mathbb{C}^{2^n}$ and a state can be written as $\ket{\phi} = \sum_{i=0}^{2^n - 1} \alpha_i \ket{i}$, where $\ket{i} = \ket{b_{n-1}}\otimes\ket{b_{n-2}}\otimes\dots\otimes\ket{b_0}$ and  the binary representation of $i$ is $b_{n-1}b_{n-2}\dots b_0$.
Additionally, $\alpha_i \in \mathbb{C}$ and $\sum_{i=0}^{2^n - 1} \abs{\alpha_i}^2 = 1$.
Quantum systems can also be combined by tensor product; the global state of an $n$-qubit and $m$-qubit system can be written as $\ket{\phi} \otimes \ket{\psi} \in \mathbb{C}^{2^n + 2^m}$, where $\ket{\phi} \in \mathbb{C}^{2^n}$ and $\ket{\psi} \in \mathbb{C}^{2^m}$.

The standard operation on an $n$-qubit system is the unitary operation, $U \in \mathbb{C}^{2^n} \cross \mathbb{C}^{2^n}$.
A unitary operation has the property that its inverse is its conjugate transpose, $U^{-1} = U^\dagger = \overline{U}^\top$.
A unitary operation, $U$, applied to a quantum state, $\ket{\phi}$, is written as $U\ket{\phi}$ and acts linearly, \ie{} $U (\sum_k \ket{\phi_k}) = \sum_k U\ket{\phi_k}$.
The application of unitaries onto a quantum state is done by dot product and written as $U_t \dots U_2 U_1 \ket{\phi}$.
Unitary operations may be performed on subsets of the system by use of the tensor product, \ie{} $(U_n \otimes U_m) (\ket{\phi} \otimes \ket{\psi}) = U_n\ket{\phi} \otimes U_m\ket{\psi}$, where $\ket{\phi} \in \mathbb{C}^{2^n}, \ket{\psi} \in \mathbb{C}^{2^m}$ and $U_n, U_m$ are respective unitary operations.


Examples of unitary operations for quantum computing are given.
The controlled Z operation, $CZ$, applies a phase of $-1$ to the state $\ket{\compone\compone}$ and does nothing otherwise.
Another is the controlled-not operation, $CNOT$, which does nothing to $\ket{\compzero\compzero}$ and $\ket{\compzero\compone}$ but changes $\ket{\compone\compzero}$ to $\ket{\compone\compone}$ and vice versa.
Their representation as unitary operations are
\begin{equation*}
\begin{aligned}
    & CZ = \begin{pmatrix}
        1 & 0 & 0 & 0 \\
        0 & 1 & 0 & 0 \\
        0 & 0 & 1 & 0 \\
        0 & 0 & 0 & -1
    \end{pmatrix} & \text{and}
    & \quad
    CNOT = \begin{pmatrix}
        1 & 0 & 0 & 0 \\
        0 & 1 & 0 & 0 \\
        0 & 0 & 0 & 1 \\
        0 & 0 & 1 & 0
    \end{pmatrix}.
\end{aligned}
\end{equation*}

A full introduction to quantum computing can be found in~\cite{Nielsen11}.

\subsection{Linear Optics}
\label{bg:optics}
\begin{figure}[t]
    \centering
    \begin{subfigure}{.45\textwidth}
        \centering
        \begin{tikzpicture}
            \pic [scale=.5] (b) {bs};
        \end{tikzpicture}
        \caption{Beam splitter}
    \end{subfigure}
    \hfill
    \begin{subfigure}{.45\textwidth}
        \centering
        \begin{tikzpicture}
            \pic (p) {ps};
        \end{tikzpicture}
        \caption{Phase shifter}
    \end{subfigure}
    \caption{Linear optical components.}
    \label{fig:bg:optics:components}
\end{figure}
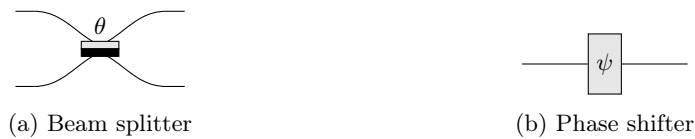

Linear optics is a means of implementing quantum computing on a physical device.
An optical circuit has a number of wires (or modes) that photons can be sent down.
Unlike quantum computing circuits, where a single qubit is sent down a single wire, a wire in an optical circuit can have several photons sent down a wire.
The base components of a linear optical circuit are beam splitters (BS), wave-plates, and phase shifters.
The base components introduce phases onto photons sent through them and can transfer photons from one wire to another (with some probability).
These components are used to construct circuits that can be used to emulate operations used by quantum computers.
The components are represented diagrammatically in Figure~\ref{fig:bg:optics:components}.

Components are represented by a transfer matrix.
For instance, a phase shift applies a shift in phase by $e^{\iu \phi}$ to a single wire, and a beam splitter acting on two wires has a transfer matrix of the form $\begin{pmatrix}
    \cos(\theta) & \iu e^{-\iu \phi} \sin(\theta)\\
    \iu e^{\iu \phi} \sin(\theta) & \cos(\theta)
\end{pmatrix}$, where $\theta$ is an angle and $\phi$ is the relative phase (usually $\phi = 0$ or $\pi$).
For a beam splitter, $\cos^2\theta$ represents the chance of being reflected back into the same wire and $\sin^2\theta$ represents the chance of a photon being transferred to the other wire and picking up some phase.
There are other components, such as wave plates, but these are not discussed.

These component operations can be combined using tensor and dot product in a similar way to quantum operations to provide a transfer matrix, $\Ulin{}$, that represents the circuit (of size $\mathbb{C}^m \times \mathbb{C}^m$ for $m$ wires).
It should be noted that the transfer matrix (i) is unitary; and (ii) can be decomposed back into components via different protocols.
For example, a pyramid of beam splitters and phase shifters can be used to decompose a transfer matrix into a circuit.
This means finding a circuit is equivalent to finding the transfer matrix.

A Fock state is a state of the wire; consisting of its phase, $\alpha \in \mathbb{C}, \abs{\alpha} \leq 1$; and the number of photons down the wire, $n \in \mathbb{Z}_+ \cup \{0\}$.
We denote the Fock state of $n$ photons and phase $\alpha$ as $\alpha \fock{n}$.
The set $\{\fock{n} : n \in  \nat \}$ acts as a basis set for a wire, \ie a wire represented by the state $\ket{\Psi}$ can be written as $\sum_{i} \alpha_i \fock{i}$, where $\alpha_i \in \mathbb{C}$ and $\sum_i \abs{\alpha_i}^2 = 1$.
Fock states can then be combined together using tensor product; for $m$ wires a Fock basis state is written as $\fock{n_1, n_2, \dots, n_m} = \fock{n_1} \otimes \fock{n_2} \dots \fock{n_m}$.

To transition between basis states on a wire, the creation and annihilation operation are used.\footnote{We do not discuss the annihilation operation, but to state simply $\hat{a}_i$ sends $\fock{n_1, \dots, n_i, \dots, n_m} \to \fock{n_1, \dots, n_i -1, \dots, n_m}$.}
The creation operator on wire $i$, $\creation{i}$, acts on the Fock state $\fock{n_1, \dots, n_m}$ by
\begin{equation*}
    \creation{i} \fock{n_1, \dots, n_i, \dots, n_m} = \sqrt{n_i + 1} \fock{n_1, \dots, n_i + 1, \dots, n_m}.
\end{equation*}
A transfer matrix, $\Ulin{}$, transforms the creation operations such that $\creation{j} \rightarrow \sum_{i=1}^{m} \Ulin{}_{ij} \creation{i}$.
Further, each transfer matrix can act on Fock states through a unitary operation, $\Ufock{}$, where
\begin{equation}
    \Ufock{} \fock{n_1, \dots, n_m} = \prod_{j=1}^{m} \frac{1}{\sqrt{n_j!}} \Big( \sum_{i=1}^{m} \Ulin_{ij} \creation{i} \Big)^{n_j} \fock{\emptyset},
    \label{eq:opt2fock}
\end{equation}
with $\fock{\emptyset} = \fock{0, \dots, 0}$ representing the vacuum Fock state (no photons through any wires).

\subsubsection{Linear Optical Quantum Computing}
Quantum computing is usually implemented on a linear optics device by using the dual rail system.
In the dual rail system, a single qubit is implemented on two wires, where the basis states, $\ket{\mathtt{0}}$ and $\ket{\mathtt{1}}$, are represented by the Fock states $\fock{1,0}, \fock{0,1}$ respectively (\ie one photon down the first wire and none down the second wire, and vice versa).
Therefore, to implement an $q$-qubit operation on the dual rail system, at least $2q$ wires are required to represent the operation.
Normally though, a quantum operation will use extra wires, known as \emph{auxiliary} wires, to implement an operation.
Thus, a quantum operation will be implemented using $2q + \numaux$ wires, where $\numaux$ is the number of auxiliary wires.

For a dual rail system with auxiliary wires, quantum computing basis states are represented by their Fock state encoding and the initial Fock states of the auxiliary wires.
For instance, a two qubit system with two \emph{vacuum} wires (no photons sent down) is represented as
\begin{equation}
\begin{aligned}
    & \ket{\compzero\compzero} = \rail{1,0,1,0}{0,0}, & \ket{\compzero\compone} = \rail{1,0,0,1}{0,0}, \\
    & \ket{\compone\compzero} = \rail{0,1,1,0}{0,0}, & \ket{\compone\compone} = \rail{0,1,0,1}{0,0}.
\end{aligned}
\label{eq:post-select-basis}
\end{equation}
The collection of such Fock states is called the coincidence basis, $\coinbasis$, and represents the desired states to be measured within a linear optics circuit.
Undesirable states can be generated by the optical circuit and these are often ignored.
In the setup given, examples of undesirable states include no photons detected in the wires for a quantum state ($\rail{1,1,0,0}{0,0}$), multiple photons in a single wire/rail ($\rail{0,0,2,0}{0,0}$), or additional photons detected in the auxiliary wires ($\rail{0,1,0,0}{1,0}$).
In general, all Fock states of $n$ photons down $m$ wires forms a basis, $\fockbasis_{n,m}$, and $\coinbasis \subset \fockbasis_{n,m}$.

There are two types of measurement that are of interest in linear optics (for quantum computing).
The first is \emph{post-selection}, where all wires are measured.
The corresponding coincidence basis contains only the basis states in quantum computing, \eg{} as given in Equation~\eqref{eq:post-select-basis}.
Any other Fock states not in the coincidence basis are ignored.

The second measurement type is \emph{heralded} selection, where photons are sent down the auxiliary wire and only the auxiliary wires are measured.
The corresponding coincidence basis contains any valid state where the auxiliary wires have the same number of photons as they started with.
Whilst quantum computing can be done with post-selection and with a higher probability of success in comparison to heralded selection; heralded is preferred over post-select because only the auxiliary wires are measured, the Fock states that represent the qubits are not measured, and therefore the quantum state can be reused to perform multiple unitary operations.

Finally, it is important to note that quantum computing operations occur probabilistically on photonic devices.
This is because the quantum computing basis (the states we are interested in) only have a chance of being measured.
Consider a transfer matrix, $\Ulin{}$, with $n$ photons and $m$ wires with the goal of simulating a $q$ qubit gate, $U$.
Let $\qubitbasis \subset \fockbasis_{n,m}$ be the set of Fock states that represent the quantum computing basis states ($\ket{b_1 \dots b_q} = \fock{n_1, \dots, n_m}$).
How $\Ulin$ acts on the basis Fock states in $\fockbasis_{n,m}$ can be seen as a matrix, $(\Ulin)_{\fockset}$, of size $\abs{\fockbasis_{n,m}} \times \abs{\fockbasis_{n,m}}$.
Then, $\Ulin$ implements $U$ on its Fock state representation if
{\newcommand{\vphB}{\vphantom{\Big\{}}
\newcommand{\addlinegap}[1][\jot]{\\\\[\dimexpr-\normalbaselineskip+#1]}
\begin{equation*}
    (\Ulin)_{\fockset} =
    \begin{array}{ r @{}}
        \abs{\qubitbasis} \Bigl\{ \addlinegap \smallskip
    \end{array}
    \left( \begin{array}{@{}c@{}}
    \vphB \addlinegap \vphB
    \end{array} \right.\kern-\nulldelimiterspace
    \overbrace{\begin{array}{@{}c@{}}
    \vphB \alpha U \addlinegap \vphB M_0
    \end{array}}^{\abs{\qubitbasis}}
    \hspace{\arraycolsep}
    \begin{array}{@{}c@{}}
    \vphB M_1 \addlinegap \vphB M_2
    \end{array}
    \kern-\nulldelimiterspace\left.\begin{array}{@{}c@{}}
    \vphB \addlinegap \vphB
    \end{array} \right),
\end{equation*}}
where $\alpha \in \mathbb{C}$, $\abs{\alpha}^2$ represents the probability of success, and $M_i$ are block matrices (whose values we do not care about).
\emph{I.e.}, the Fock operation of $\Ulin$ acts like $U$ on the quantum computing basis, $\qubitbasis$, up to some factor, $\alpha$.
More details are provided in Section~\ref{sec:method:encoding}.

\begin{figure}[t!]
    \centering
    \begin{subfigure}{.35\textwidth}
        \centering
        \scalebox{.8}{\begin{tikzpicture}
    \begin{pgfonlayer}{nodelayer}
        \node (p0l) at (-2, 3.5) {};
        \node (p0r) at (2, 3.5) {};
        \node (p1l) at (-2, 2.5) {};
        \node (p1r) at (2, 2.5) {};
        \node (p2l) at (-2, 2) {};
        \node (p2r) at (2, 2) {};
        \node (p3l) at (-2, 1) {};
        \node (p3r) at (2, 1) {};
        \node (p4l) at (-2, .5) {};
        \node (p4r) at (2, .5) {};
        \node (p5l) at (-2, -.5) {};
        \node (p5r) at (2, -.5) {};
        \pic [scale=.5] (bs1) at (0,3) {bs};
        \pic [scale=.5] (bs2) at (0,1.5) {bs};
        \pic [scale=.5] (bs3) at (0,0) {bsr};
    \end{pgfonlayer}
    \begin{pgfonlayer}{edgelayer}
        \draw (p0l.center) to (-1.125,3.5);
        \draw (1.125,3.5) to (p0r.center);
        \draw[in=180, out=0] (p1l.center) to (-1.125,2);
        \draw[in=-180, out=0] (1.125,2) to (p1r.center);
        \draw[in=180, out=0] (p2l.center) to (-1.125,1);
        \draw[in=180, out=0] (1.125,1) to (p2r.center);
        \draw[in=180, out=0] (p3l.center) to (-1.125,.5);
        \draw[in=180, out=0] (1.125,.5) to (p3r.center);
        \draw[in=180, out=0] (p4l.center) to (-1.125,-.5);
        \draw[in=180, out=0] (1.125,-.5) to (p4r.center);
        \draw[in=180, out=0] (p5l.center) to (-1.125,2.5);
        \draw[in=180, out=0] (1.125,2.5) to (p5r.center);
    \end{pgfonlayer}
\end{tikzpicture}}
        \caption{Post-select scheme; $\theta$ is such that $\sin^2\theta = 1/3$~\cite{Ralph02}.}
    \end{subfigure}
    ~
    \begin{subfigure}{.6\textwidth}
        \centering
        \scalebox{.8}{\tikzset{
    bsa/.pic={\bsset{$\theta_1$}},
    bsb/.pic={\bsset{$\theta_2$}},
    bsc/.pic={\bsset{$\theta_3$}},
}

\scalebox{1}{
\begin{tikzpicture}
    \begin{pgfonlayer}{nodelayer}
        \node (p0l) at (-4, 3.25) {};
        \node (p0r) at (4, 3.25) {};
        \node (p1l) at (-4, 2.5) {};
        \node (p1r) at (4, 2.5) {};
        \node (p2l) at (-4, 2) {};
        \node (p2r) at (4, 2) {};
        \node (p3l) at (-4, 1) {};
        \node (p3r) at (4, 1) {};
        \node (p4l) at (-4, .5) {};
        \node (p4r) at (4, .5) {};
        \node (p5l) at (-4, -.5) {};
        \node (p5r) at (4, -.5) {};
        \pic [scale=.5] (ps1) at (-3, 2.5) {ps};
        \pic [scale=.5](ps5) at (-3, -.5) {ps};
        \pic [scale=.5] (bs2) at (-1,1.5) {bsa};
        \pic [scale=.5] (bs3) at (2, 1.5) {bsb};
        \pic [scale=.5] (bs2) at (-1,0) {bsa};
        \pic [scale=.5] (bs3) at (2, 0) {bsc};
    \end{pgfonlayer}
    \begin{pgfonlayer}{edgelayer}
        \draw (p0l.center) to (p0r.center);
        
        \draw (p1l.center) to (-3.5,2.5);
        \draw[in=180, out=0] (-2.5,2.5) to (-2.1,2);
        \draw (0.1,2) to (.9,2);
        \draw[in=180, out=0] (3.1,2) to (3.5,2.5);
        \draw (3.5,2.5) to (p1r.center);
        
        \draw (p2l.center) to (-2.5,2);
        \draw[in=180, out=0] (-2.5, 2) to (-2.1, 2.5);
        \draw (-2.1, 2.5) to (3.1, 2.5);
        \draw[in=180, out=0] (3.1, 2.5) to (3.5,2);
        \draw (3.5,2) to (p2r.center);
        
        \draw (p3l.center) to (-2.5,1);
        \draw[in=180, out=0] (-2.5, 1) to (-2.1, .5);
        \draw[in=180, out=0] (.1, .5) to (.9, 1);
        \draw (3.1,1) to (p3r.center);

        \draw (p4l.center) to (-2.5,.5);
        \draw[in=180, out=0] (-2.5, .5) to (-2.1, 1);
        \draw[in=180, out=0] (.1, 1) to (.9, .5);
        \draw (3.1,.5) to (p4r.center);

        \draw (p5l.center) to (-3.5,-.5);
        \draw (-2.5,-.5) to (-2.1,-.5);
        \draw (0.1,-.5) to (.9,-.5);
        \draw (3.1,-.5) to (p5r.center);
        \end{pgfonlayer}
\end{tikzpicture}
}}
        \caption{Heralded scheme; $\psi = \pi$ and $\theta_i$ ($i=1,2,3$) are radian values equivalent to those in \cite{Knill02}.}
    \end{subfigure}
    \caption{Optical circuits of the controlled-$Z$ operation.
    The first two wires act as the control qubit, the middle two as the target qubit, and the last two are ancillary wires.}
    \label{fig:bg:optics:cz}
\end{figure}
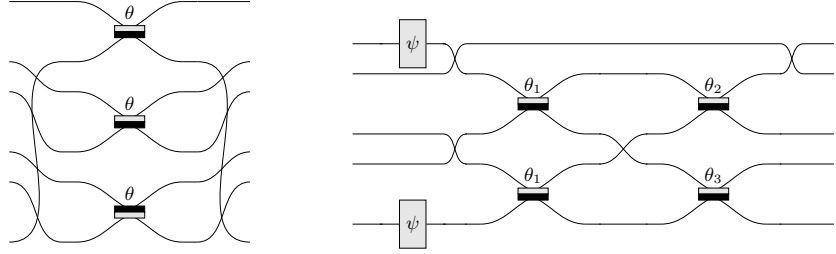

We now give examples of quantum computing operations implemented in the post-select and heralded schemes.
In Figure~\ref{fig:bg:optics:cz}, the photonics circuits of the $CZ$ operation is shown using post-selection and heralded selection.
In the post-selected case, two vacuum ancillary wires (no photons sent down) are used to obtain the transfer matrix
\begin{equation}
    \frac{1}{3}\begin{pmatrix}
    \sqrt{3} & 0 & 0 & 0 & 0 & -\sqrt{6} \\
    0 & \sqrt{3} & 0 & -\sqrt{6} & 0 & 0 \\
    0 & 0 & \sqrt{3} & 0 & \sqrt{6} & 0 \\
    0 & \sqrt{6} & 0 & \sqrt{3} & 0 & 0 \\
    0 & 0 & -\sqrt{6} & 0 & \sqrt{3} & 0 \\
    -\sqrt{6} & 0 & 0 & 0 & 0 & -\sqrt{3}
    \end{pmatrix},
    \label{eq:optic:cz}
\end{equation}
which has a probability of success of $\frac{1}{9}$~\cite{Ralph02}.

In the heralded case, the circuit uses two ancillary wires with a photon sent down each ancillary wire and has a probability of success of $\frac{2}{27}$~\cite{Knill02}.
The associated transfer matrix is
\begin{equation*}
    \begin{pmatrix}
    1 & 0 & 0 & 0 & 0 & 0 \\
    0 & -\frac{1}{3} & 0 & -\frac{\sqrt{2}}{3} & \frac{\sqrt{2}}{3} & \frac{2}{3} \\
    0 & 0 & 1 & 0 & 0 & 0 \\
    0 & \frac{\sqrt{2}}{3} & 0 & -\frac{1}{3} & -\frac{2}{3} & \frac{\sqrt{2}}{3} \\
    0 & -\frac{\sqrt{3-\sqrt{6}}}{3} & 0 & \frac{\sqrt{3-\sqrt{6}}}{3} & -\frac{\sqrt{3 + \sqrt{6}}}{3\sqrt{2}} & \sqrt{\frac{1}{6} - \frac{1}{3\sqrt{6}}} \\
    0 & -\frac{\sqrt{3-\sqrt{6}}}{3} & 0 & -\frac{\sqrt{3-\sqrt{6}}}{3} & -\sqrt{\frac{1}{6} - \frac{1}{3\sqrt{6}}} & -\frac{\sqrt{3 + \sqrt{6}}}{3\sqrt{2}}
    \end{pmatrix}.
\end{equation*}

\subsection{The Problem}
\label{bg:problem}
To summarise, a photonics circuit of $m$ wires can be represented by a $m \times m$ transfer matrix, $\Ulin$.
Through a procedure, this can be transformed into an operation, $\Ulin_{\fockset}$, that works on a larger space, where we wish to ensure a particular subspace of the operation acts like a desired quantum computing operation, $U$, up to some factor $\alpha \in \mathbb{C}$.
There is only a chance of being in this subspace, and this probability of success is $\abs{\alpha}^2$.

The simplest problem to consider is the verification version of the problem.
\begin{problem}
    Given a transfer matrix, $\Ulin$, check that a quantum computing operation, $U$, is implemented on a subspace of the Fock space, $\Ulin_{\fockset}$, up to some factor, $\alpha$.
\end{problem}
The harder problem we are considering is the synthesis of the transfer matrix.
\begin{problem}
Given a quantum computing unitary operation, $U$, that acts on $q$ qubits, find a photonics circuit, represented by $\Ulin$, that implements $U$ on its Fock space $\Ulin_{\fockset}$.
\end{problem}
In this general setting for the problem, one can consider various setups and different ways to encode quantum bits into Fock states.
We restrict this problem into working on a specific basis and using the dual-rail encoding, where a qubit in the quantum computing space is represented by two optical wires.
Additionally, it is important to maximize the likelihood of the operation is performed successfully.
Thus, the problem needs to be optimized.
This leaves us with the following problem.
\begin{problem}
    \label{prob:linopt}
    Given a quantum computing unitary operation that acts on $q$ qubits, $U$, $2q + \numaux$ optical wires, and a coincidence basis, $\coinbasis$; find the transfer matrix of a dual-rail optical operation, $\Ulin$, that maximises the likelihood of $U$ occurring.
\end{problem}

\section{Method}
\label{sec:method}

\subsection{Non-linear Real Arithmetic (NRA) and \texorpdfstring{$\delta$}~-Satisfiability}
\label{sec:method:dsat}
Generally, SMT solver theories are usually $\NP$-hard but are decidable, both LIA and LRA are as such (linear integer/real arithmetic respectively).
Some theories though are undecidable.
One such theory, Non-linear Real Arithmetic (NRA), is the theory that will be used for solving our problem.
Standard NRA mainly consists of multivariable polynomials, but can be extended to consist of non-linear functions; such as powers of variables, and trigonometric functions.

However, it was shown in~\cite{deltasat} that by allowing a small perturbation in NRA expressions, the theory can become decidable.
This perturbation comes in the form of a value $\delta \in \posrat$, which is used to weaken arithmetic expressions, $f(x) = 0$, such that they are of the form $\abs{f(x)} \leq \delta$.
This process is known as $\delta$-weakening.
Instead of returning the standard \sat{} or \unsat{}, an SMT solver that implements $\delta$-weakening will instead return \txtdsat{}, where the $\delta$-weakening of the expressions is satisfiable, or \unsat{}, where the expressions are unsatisfiable.

The tool dReal~\cite{dreal} implements $\delta$-weakening and, when \txtdsat{}, returns a region of values that satisfy the $\delta$-weakened expressions.
This enables it to find $\delta$-satisfiability for (an extension of) NRA expressions.
We write
\begin{equation*}
    sat, regions \gets \texttt{dReal}( c, \delta),
\end{equation*}
to mean dReal, run with the constraints $c$ and precision $\delta$, returns \txtdsat{}, \unsat{}, or \unkown{} (in the case of timeouts or crashes); and, if dReal returns \txtdsat{}, return the satisfiable regions or $[~]$ otherwise.
For example, if $\delta = 0.001$, the expressions
\begin{equation*}
    (x + 2^y = 3) \land (y > \sin(x)) \land (x > 0.5) 
\end{equation*}
is \txtdsat{} with $x \in [0.8685\dots, 0.8687\dots]$ and $y \in [1.0916\dots, 1.0918\dots]$.
This tool forms the basis of how transfer matrices in the coincidence basis are found.

Whilst NRA is solvable using standard SMT solvers, we found that using dReal would produce a region of approximations when standard tools could not find an exact solution or dReal would find a region much faster.
However, this means that instead of getting a transfer matrix that can be used, we have a region that approximately solves our constraints.
In Section~\ref{sec:method:unitary_approx}, we describe how we can find an actual transfer matrix from our region of approximations, allowing us to return an exact solution.

\subsection{Encoding of Photonics Problem into NRA}
\label{sec:method:encoding}
Given Problem~\ref{prob:linopt}, we show how we can transform the problem into a problem in non-linear real arithmetic (NRA).
As a reminder, we are given a quantum computing operation, $U$, and a coincidence basis, $\coinbasis$; and need to find a suitable $m \times m$ transfer matrix, $\Ulin$, that implements $U$ with a probability of success, $\abs{\probamp}^2$, where $m$ is the number of wires and $\probamp$ denotes the success amplitude.
Thus, we consider $\Ulin_{ij}$ and $\probamp$ to be real variables.
\begin{remark}
In general, it can be that $\probamp, \Ulin_{ij} \in \mathbb{C}$.
For ease of demonstrating the process of converting into NRA, we restrict $\Ulin$ to be a real matrix and $\probamp \in \mathbb{R}$.
However, it is possible to encode variables as two real variables consisting of the real and imaginary part, and modify the following encoding for the complex representation.
We discuss the consequences of allowing variables to be complex in Section~\ref{sec:res:discussion}.
\end{remark}

To begin with, we know that $\Ulin$ needs to be unitary, \ie{} $\Ulin \Ulin^\dagger{} = I$.
This can be encoded as
\begin{equation*}
    \mathtt{unitary} = \bigwedge_{1\leq i \leq m} \Big(\sum_{1 \leq k \leq m} \Ulin_{ik} \Ulin^\dagger_{ki} = 1 \Big)
    \land
    \bigwedge_{1\leq i\neq j \leq m} \Big(\sum_{1 \leq k \leq m} \Ulin_{ik} \Ulin^\dagger_{kj} = 0 \Big).
\end{equation*}

Additionally, since $\Ulin$ is unitary, we can restrict the values that $\Ulin$ can take to be between $-1$ and $1$.
Further, the probability of success ($\abs{\probamp}^2$) cannot be greater than $1$.
Thus, these can simply be encoded as
\begin{equation*}
    \mathtt{bound} = (-1 \leq \probamp) \land (\probamp \leq 1) \land \Big( \bigwedge_{1 \leq i, j \leq m} -1 \leq \Ulin_{ij} \leq 1 \Big).
\end{equation*}

The core constraint to be considered is that $\Ulin$ implements $U$ on its Fock matrix $(\Ulin)_\fockset$ up to the factor $\probamp$.
Our coincidence basis consists of the set of Fock states for the quantum computational basis states, $\qubitbasis$; and, if heralded, the set containing any other Fock state with the same number of photons in each auxiliary wires, $\coinbasis_{h}$, \ie{} $\coinbasis = \qubitbasis$ if using the post-select regime and $\coinbasis = \qubitbasis \cup \coinbasis_h$ if using the heralded regime.
For $\fock{\phi} \in \qubitbasis$, let $\ket{b}$ be the quantum computational representation of $\fock{\phi}$, \eg for a vacuum auxiliary of two wires, $\ket{\compzero \compzero}$ is the representation of $\fock{\phi} = \fock{1010}\fock{00}$.
Then, the Fock states acting on the encoded quantum computational basis states are required to be equal up to the probability amplitude, \ie{}
\begin{equation*}
    \mathtt{fockequal} = \bigwedge_{\ket{b}\cong\fock{\phi} \in \coinbasis_{qc}} \bigwedge_{\ket{c}\cong\fock{\psi} \in \coinbasis_{qc}} (\probamp U_{bc} = (\Ulin_\fockset)_{\phi\psi} ).
\end{equation*}

Thus, the simplest encoding of Problem~\ref{prob:linopt} in NRA is
\begin{equation*}
    \mathtt{core} = \mathtt{bound} \land \mathtt{unitary} \land \mathtt{fockequal}.
\end{equation*}
with $\mathtt{bound}$, $\mathtt{unitary}$, and $\mathtt{fockequal}$ having $m^2 + 2$, $m^2$, and $(2^q)^2$ formulae respectively, where $m$ is the number of modes and $q$ is the number of qubits to simulate.

\begin{remark}
    To represent a $q$-qubit operation, $k = q + n_a$ photons are used, where $n_a$ is the number of photons going down the auxiliary wires.
    The constraints given in $\mathtt{core}$ are a combination of $2$- and $k$-degree polynomials.
    The $2$-degree polynomials come from the $\mathtt{unitary}$ constraint and the polynomials from the $\mathtt{fockequal}$ constraints are $k$-degree since we are using the Fock version of $\Ulin$ (see Equation~\ref{eq:opt2fock}).
    The number of photons down each auxiliary wires additionally affects the coefficients of the polynomials.
\label{rmk:photondeg}
\end{remark}

However, we can also include additional constraints based on the unitary gate required or the coincidence basis being used.
For instance, for a $CZ$ operation, we can set the first and third wire to not interact with other wires representing the qubits, \ie{} $\Ulin_{1j}$ and $\Ulin_{j1}$ are set to be equal to $0$ for $j \in \{2,3,4\}$ and, similarly, $\Ulin_{3k}, \Ulin_{k3}$ for $k \in \{1,2,4\}$.
This can be done since the $CZ$ operation does nothing when the control qubit is in the $\ket{\compzero}$ and, even when the state is controlled, there is no interaction with the target qubit when it is in the $\ket{\compzero}$ state.
This means that the second and fourth wires are the only wires that interact with auxiliary wires (if there are any).


Alternatively, we can also enforce constraints on the auxiliary wires, particularly vacuum auxiliary wires.
We can set it such that a vacuum wire need not be used (\emph{vacuum relaxing}, $\Ulin_{kk} = 1 \lor \Ulin_{kk} < 1$) or that a vacuum wire must be used (\emph{vacuum enforcing}, $\Ulin_{kk} \Ulin^\dagger_{kk} \neq 1$).
Vacuum relaxing allows the SMT solver to easily consider cases when the vacuum wire may not be used (meaning a smaller circuit can be considered) and vacuum enforcing requires the resulting circuit to use the vacuum wire in some way.

\subsection{Searching using dReal}
\label{sec:method:search}
\begin{algorithm}[t]
    \caption{Search Algorithm using dReal}
    \label{alg:search}
    \SetKwInOut{Input}{Input}
    \Input{unitary operation, $U$; coincidence basis, $\coinbasis$; minimum threshold, $\probamp_{min} \geq 0$; precision, $\delta > 0$; timeout $ > 0$.}
    result $\gets$ \unchecked\;
    regions $\gets [~]$\;
    sat-result $\gets \dsat$\;
    $ \mathtt{consts} \gets constraints(U, \coinbasis)$ \;
    \While{$sat-result = \dsat \land runtime \leq timeout$}{
        constraints $\gets \mathtt{consts} \land (\abs{\probamp}^2 \geq \probamp_{min})$\;
        sat-result, regions $\gets$ \texttt{dReal}$(\text{constraints}, \delta)$\;
        \If{sat-result $= \dsat$}{
            $\probamp_{min} \gets \abs{\ub{\text{regions}[\probamp]}}^2 + \frac{1}{10} \delta$\;
            result $\gets$ \approximate{}\;
        }
    }
    \lIf{result $= \approximate~\land$ sat-result $= \unsat$}{result $\gets$ \doptimal}
    \lIf{result $= \unchecked~\land$ sat-result $= \unsat$}{result $\gets \infeasible$}
    \lIf{result $= \unchecked~\land$ sat-result $= \unkown$}{result $\gets \unkown$}
    \Return result, regions
\end{algorithm}
The algorithm for how we can synthesise and find an optimal transfer matrix using dReal is described in Algorithm~\ref{alg:search}.
We have shown how to make constraints for the operation we are synthesizing given a quantum computing unitary matrix, $U$, and coincidence basis, $\coinbasis$, in Section~\ref{sec:method:encoding}.
Thus, we have
\begin{equation*}
    constraints(U, \coinbasis) = \mathtt{core} \land \mathtt{extra},
\end{equation*}
where $\Ulin$ is represented by real variables and $\mathtt{extra}$ encodes any extra constraints required by the user (constraints on gates, vacuum relaxing, \etc{}).
In practice, the constraints are generated in standard SMT-LIB format~\cite{smt2lib}.
The \texttt{dReal} function used in the algorithm is described in Section~\ref{sec:method:dsat}.

The algorithm starts with an initial probability of success to beat and tries to find a transfer matrix that has a higher probability of success.
If it is able to find one, then it updates the value the probability of success needs to beat by going slightly above the upper bound of the probability of success that was found ($\probamp_{min} \gets \abs{\ub{\text{regions}[\probamp]}}^2 + \frac{1}{10} \delta$), and tries to search again.
It continues polling for a set time limit or if a call returns \unsat{} before stopping.

The algorithm returns one of four possible outputs:
\begin{itemize}
    \item \doptimal: a valid region was found and it is optimal up to $\delta$ (\ie the probability of success cannot get higher);
    \item \approximate: a valid region was found (but it might not be optimal);
    \item \unkown: no region was found within the time limit (but one may exist);
    \item \infeasible: there is no valid transfer matrix with probability amplitude greater than the initial $\probamp_{min}$.
\end{itemize}
If the output is \doptimal{} or \approximate{}, then the algorithm also returns the valid regions for the variables.
From this, we can get out a region for $\Ulin$ and a region for the probability amplitude $\probamp$.

\begin{remark}
    It is possible to use standard SMT solvers (\eg Z3~\cite{Z3Solver}, Yices~\cite{Yices2}, cvc5~\cite{cvc5}) instead of dReal, which would lead the algorithm to find an exact solution.
    Algorithm~\ref{alg:search} can be modified to call the appropriate solver and replace \doptimal{} with \texttt{optimal}.
    However in practice, we found that beyond simple examples, dReal would be capable of finding an approximation whereas other solvers would take much longer or could not find ome.
\end{remark}

\subsection{Finding a Unitary from Approximations}
\label{sec:method:unitary_approx}
If the search from Section~\ref{sec:method:search} was successful, two matrices will be returned, $\lb{\Ulin}$ and $\ub{\Ulin}$, where for any matrix $A$ that has elements $(\lb{\Ulin})_{i j} \leq (A)_{i j} \leq (\ub{\Ulin})_{i j}$, $A$ satisfies the $\delta$-weakened expressions.
However, these matrices are not unitary.
This is because $A$ satisfies the $\delta$-weakened expressions and so $A A^\dagger = I + \omega$, where $\abs{\omega_{ij}} \leq \delta$.
Therefore, a suitable unitary matrix needs to be found around the region contained by $\lb{\Ulin}$ and $\ub{\Ulin}$ to be the final returned transfer matrix.

A unitary matrix that is close this region can be found by taking choosing a matrix within the region, $A$, and considering its singular value decomposition, $A = V \Sigma W$~\cite{Reich,UnitaryBlog}.
Here, $V$ and $W$ are unitary matrices and $\Sigma$ is a diagonal matrix.
Since $A$ is close to being unitary, the elements of the diagonal matrix are close to 1, \ie $(\Sigma)_{ii} \approx 1$.
By replacing $\Sigma$ with the identity matrix and since $V$ and $W$ are unitary matrices, then $\hat{A} = VIW = VW$ is a unitary matrix that is close to our original matrix $A$.

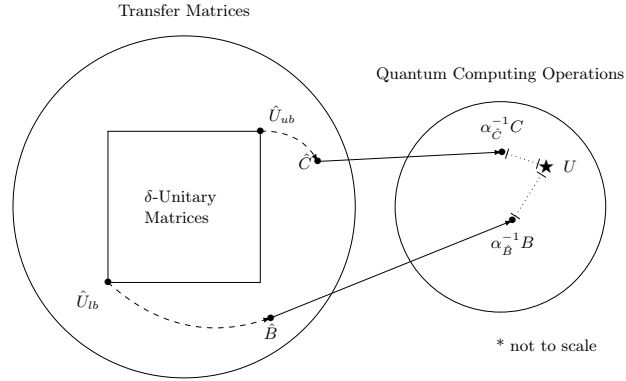
\begin{figure}[t]
    \centering
    \scalebox{.7}{\begin{tikzpicture}[
    label distance=-.2cm,
    every edge/.style = {draw, -latex, dashed},
    ]
    \begin{pgfonlayer}{nodelayer}
        \node[label={below left:$\lb{\Ulin}$}] (Ulb) [] {$\bullet$};
        \node[label={above right:$\ub{\Ulin}$}] (Uub) [above right=2.5cm and 2.5cm of Ulb] {$\bullet$};
        \coordinate (lincentre) at ($(Ulb)!0.5!(Uub)$);
        \node[draw, shape=circle,minimum size=6.5cm,label={[label distance=.25cm]above:{Transfer Matrices}}] (linspace) at (lincentre) {};
        
        \node[label={below:$\hat{B}$}] (Blin) [below right = .3cm and 2.7cm of Ulb] {$\bullet$};
        \node[label={left:$\hat{C}$}] (Clin) [below right = .2cm and .7cm of Uub] {$\bullet$};
        
        \node[draw,shape=circle,minimum size=4cm, label={[label distance=.25cm]above:{Quantum Computing Operations}}] (qcspace) [right = 4cm of lincentre] {};
        \node[inner sep=0cm,label={[label distance=.05cm]right:$U$}] (U) [above right = -.8cm and -.7cm of qcspace] {$\bigstar$};
        \node[inner sep=0cm,label={[label distance=.05cm]below:$\probamp_{\hat{B}}^{-1} B$}] (B) [below left = .8cm and .4cm of U] {$\bullet$};
        \node[inner sep=0cm,label={[label distance=.05cm]above:$\probamp_{\hat{C}}^{-1} C$}] (C) [above left = .05cm and .6cm of U] {$\bullet$};

        \node[] [below = 3cm of U] {\small{* not to scale}};
    \end{pgfonlayer}
    \begin{pgfonlayer}{edgelayer}
        \draw[black] (Ulb) rectangle ++(Uub) node[pos=.5, text width=1.5cm] {\small{$\delta$-Unitary\\Matrices}};
        \draw[bend right] (Ulb.center) edge (Blin.center);
        \draw[bend left] (Uub.center) edge (Clin.center);
        \draw[-latex] (Blin.center) -- (B.center);
        \draw[-latex] (Clin.center) -- (C.center);
        \draw[dotted,|-|] (C) -- (U);
        \draw[dotted,|-|] (B) -- (U);
    \end{pgfonlayer}
    \end{tikzpicture}}
    \caption{Visual explanation for obtaining unitary from approximation.}
    \label{fig:unitary-approx}
\end{figure}

We apply the singular value decomposition to $\lb{\Ulin}$ and $\ub{\Ulin}$ to get two transfer matrices, $\hat{B}$ and $\hat{C}$ respectively, and then check to see which one more accurately describes our desired quantum computing operation $U$.
To decide between the matrices, we take the part of the matrix that acts on the coincidence basis of $\hat{B}$ and $\hat{C}$ ($B$ and $C$ respectively); find their probability amplitude and scale the matrices by the inverse ($\probamp_{\hat{B}}^{-1} B, \probamp_{\hat{C}}^{-1} C$); and then compare the Frobenius norm against the desired unitary matrix ($\frob{U - \probamp_{\hat{B}}^{-1} B}$, $\frob{U - \probamp_{\hat{C}}^{-1} C}$).
Whichever norm is smaller, the appropriate matrix is chosen.
Figure~\ref{fig:unitary-approx} gives a visualisation.

Thus the final algorithm is relatively simple:
\begin{enumerate}
    \item Perform the search as described in Algorithm~\ref{alg:search} with the appropriate inputs;
    \item If an approximated region was found, perform the SVD technique described on the region and return a valid transfer matrix.
    If no region was found, return $\unsat$/$\unkown$.
\end{enumerate}

\section{Results}
\label{sec:results}
The techniques described are implemented in a tool.
We validate our tool using both positive and negative known results in the literature, and then explore previously unknown results.
We divide this section based on whether the coincidence basis is post-selected (Section~\ref{sec:res:post}) or heralded (Section~\ref{sec:res:heralded}).
The commands used to provide different results are provided in the repository.

All experiments are performed on a laptop with an Intel(R) Core(TM) Ultra 7 165H \@ 4.30 GHz $\times$ 16 cores processor and 32 GB of RAM using Ubuntu 24.04.3 LTS.
All experiments are available on Zenodo.\footnote{Available at \url{https://doi.org/10.5281/zenodo.17116446}}

\subsection{Post-Selection}
\label{sec:res:post}
\subsubsection{Standard Known Results}\label{sec:std-known-postselect}
One of the main gates of importance for linear optics to implement is the $CZ$ gate.
With this gate, one can easily construct any other controlled-$U$ operation, where $U$ is a single qubit operation, by wrapping the $CZ$ gate with appropriate single unitary operations on the target qubit.
For instance, a controlled-not operation, $CNOT$, can be decomposed into the operations $(I \otimes H) CZ (I \otimes H)$, where $H$ is the Hadamard operation ($H\ket{\compzero} = \frac{\ket{\compzero} + \ket{\compone}}{\sqrt{2}}$, $H\ket{\compone} = \frac{\ket{\compzero} - \ket{\compone}}{\sqrt{2}}$).

With our tool we can verify a few known facts about the $CZ$ operation:
\begin{itemize}
    \item a $CZ$ operation is \infeasible{} with no auxiliary wires;
    \item a $CZ$ operation is \infeasible{} with a single auxiliary vacuum wire;
    \item a $CZ$ operation is found with two auxiliary vacuum wires and has the same probability of success as the transfer matrix given in Equation~\eqref{eq:optic:cz} as $\delta$ approaches 0;\footnote{Modulo phase changes.}
    \item and the above results hold for the $CNOT$ operation as well.
\end{itemize}

These checks can be performed in a matter of seconds.
Thus, for post-selection, our tool is capable of producing known results.

\subsubsection{Different Auxiliary Wire Setups}
\label{sec:res:post:onewire}
It is important to explore what other auxiliary wire setups can be used to synthesise circuits with different probabilities of success to see if higher success rates are possible.
Additionally, being able to synthesize different circuits with different auxiliary setups may demonstrate certain behaviours as limits are reached.
In this section, we investigate what happens when a single auxiliary wire is used but a different number of photons are sent down the wire.
This has been investigated in \cite{Alessio24}.

\begin{figure}[t]
    \centering
    \includegraphics[width=0.7\linewidth]{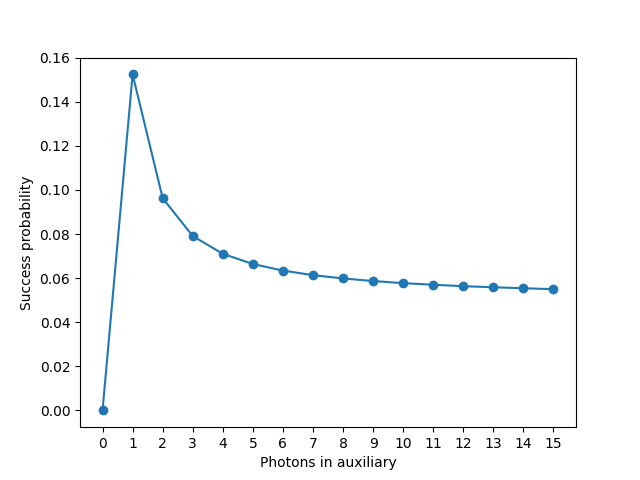}
    \caption{Best found success probability for a $CZ$ operation by sending numerous photons down one auxiliary wire.}
    \label{fig:czoneaux}
\end{figure}

The results for synthesizing a $CZ$ operation by sending different number of photons down a single auxiliary wire are visualized in Figure~\ref{fig:czoneaux}.
Our results are similar to those made in \cite{Alessio24} (see Section 3 and Appendix D therein) and seem to suggest that these are close to the optimal success probabilities (for optical circuits with real phases).
For example, our tool can show that is is \infeasible{} for a single auxiliary wire with one photon sent down to have a success probability higher than $0.16$ using only real values.

As can be seen in the results, initially a single photon sent down a single wire has a higher probability than using two vacuum wires.
However, as the number of photons increase, the probability begins to decrease.
There is a drop in probability to begin with, but the the success rate begins to decrease at a lower rate as the number of photons increases.
It is an open question on whether this converges towards $0$ or some other value.

\begin{remark}
    The polynomials generated within the constraints will have degree $2 + n_a$, where $n_a$ is the number of photons in the auxiliary wire ($q = 2$ for $q$ in Remark~\ref{rmk:photondeg}).
    However, most of the terms have a single variable of high degree since that is where most of the photons originate from and must go, \ie the terms of the polynomials have one factor of the form ${U_{ij}}^d$ where $d \geq n_a - 2$.
    This and the fact that the polynomials have very few terms, due to the coincidence basis requiring a large number of photons in the auxiliary wire, is why these high degree polynomials are solvable in a short time using dReal.
\end{remark}

\subsubsection{Givens Rotation Gates}
\label{sec:res:post:givens}
A Givens rotation operation is an unitary operation that performs an entanglement on the logical basis states $\ket{\compzero\compone}$ and $\ket{\compone\compzero}$ and does nothing to the $\ket{\compzero\compzero}$ and $\ket{\compone\compone}$ basis states.
The Givens rotation operations form a universal set of operations for problems within quantum chemistry \cite{Arrazola2022universalgivens}.
The family of Givens rotation operations that we investigate are of the form
\begin{equation}
    G(\theta) = \begin{pmatrix}
        1 & 0 & 0 & 0 \\
        0 & \cos(\theta) & -\sin(\theta) & 0 \\
        0 & \sin(\theta) & \cos(\theta) & 0 \\
        0 & 0 & 0 & 1
    \end{pmatrix},
\end{equation}
with a few examples provided:
\begin{align*}
    & G(\frac{\pi}{4}) = \begin{pmatrix}
        1 & 0 & 0 & 0 \\
        0 & \frac{1}{\sqrt{2}} & -\frac{1}{\sqrt{2}} & 0 \\
        0 & \frac{1}{\sqrt{2}} & \frac{1}{\sqrt{2}} & 0 \\
        0 & 0 & 0 & 1
    \end{pmatrix},
    & G(\frac{\pi}{2}) = \begin{pmatrix}
        1 & 0 & 0 & 0 \\
        0 & 0 & -1 & 0 \\
        0 & 1 & 0 & 0 \\
        0 & 0 & 0 & 1
    \end{pmatrix}.
\end{align*}
Variations with complex factors in the entangling and non-entangling parts do exist, \eg one variation sends $\ket{\compone\compone}$ to $e^{i\phi}\ket{\compone\compone}$, where $\phi \in [0,2\pi)$ is a parameter.

Our tool is capable of modelling and finding transfer matrices for Givens rotations.
For example, with $\theta = \frac{\pi}{2}$, we have the following transfer matrix that uses two vacuum auxiliary wires,
\begin{equation*}
    \frac{1}{3}
    \begin{pmatrix}
        0 & 0 & -\sqrt{3} & 0 & 0 & -\sqrt{6}\\
        \sqrt{6} & 0 & 0 & -\sqrt{3} & 0 & 0\\
        \sqrt{3} & 0 & 0 & \sqrt{6} & 0 & 0\\
        0 & \sqrt{3} & 0 & 0 & \sqrt{6} & 0\\
        0 & -\sqrt{6} & 0 & 0 & \sqrt{3} & 0\\
        0 & 0 & \sqrt{6} & 0 & 0 & -\sqrt{3}
    \end{pmatrix},
\end{equation*}
and it succeeds with a probability of $\frac{1}{9}$.
It looks similar to a CZ gate because the operation itself is simply a permuted CZ gate.

\begin{figure}[t]
    \centering
    \includegraphics[width=0.7\linewidth]{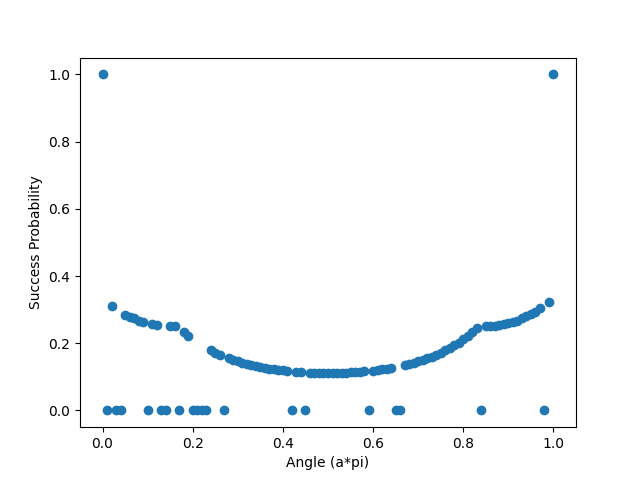}
    \caption{Plot of found success probability for Givens operations with different angles using two vacuum wires, and a single run with a 180 second timeout.}
    \label{fig:givens}
\end{figure}

The trend for Givens rotation gates against their found probability of success is mapped out in Figure~\ref{fig:givens} with angles between 0 and $\pi$.\footnote{We only investigate values of $\theta$ between $0$ and $\pi$ since $G(\theta)^{-1} = G(-\theta)$ (as $\cos(\theta) = \cos(-\theta)$ and $\sin(\theta) = - \sin(-\theta)$).
For $-\pi \leq -\theta \leq 0$, the circuit of $G(-\theta)$ is found by inverting the circuit of $G(\theta)$.}
About one quarter of the results couldn't find a solution in the time (those with zero success probability), however running more instance or for a longer time would reveal a success probability in line with the rest of the data.

The circuits for angles $0$ and $1$ are exact since the related matrices are the identity and the $Z$ gate applied to each qubit respectively (both exactly implementable).
The minima of $\frac{1}{9}$ can be observed when the angle is close to $0.5$.
The success probability initially drops substantially but then drops at a lower rate.

It can be seen that there is some relation between the probability of success of a Givens rotation gate to its angle.
There exists a similar relation for controlled phase gates~\cite{Kieling10}, where there is a formula for the optimum probability of success given the angle of the phase ($e^{\iu\phi}$), \ie the optimal success probability at angle $\phi$ is determined by a function $f(\phi)$.
However, the controlled phase gate observes non-monotone structure and has a minima ($\min_{\phi} f(\phi)$) when $\phi$ is between $\pi/3$ and $\pi$.
The Givens gates may also have a non-monotonous structure for $0 \leq \theta \leq \frac{\pi}{2}$ depending on the formula for determining the optimal success probability parametrized by $\theta$ ($g(\theta)$), but the data gathered seems to indicate $\pi/2$ exactly being the minima ($\min_{\theta} g(\theta)$).

\subsection{Heralded: Replication of Known Results}
\label{sec:res:heralded}
Now we look at some results for the heralded scheme.
With the tool, we are able to prove infeasibility of some trivial and known results:
\begin{itemize}
    \item a $CZ$ operation is \infeasible{} with no or one vacuum wire, and two vacuum wires is \infeasible{} with probability at least $\frac{1}{100}$;
    \item a $CZ$ operation is \infeasible{} with a single photon down one auxiliary wire.\footnote{Further tests suggest a single wire is \infeasible{} for any number of photons.}
\end{itemize}
Besides these results of impossibility of certain setups, we have found that our technique returns \unkown{} (due to timing out) when trying to find a known instance of a $CZ$ operation, \eg two ancillary wires with one photon down each, which has a known implementation.
This is likely due to the increased degree of the polynomials and the increase in the number of equations required to solve.




\subsection{Discussion}
\label{sec:res:discussion}
As can be seen with the Givens examples in Section~\ref{sec:res:post:givens}, photonic quantum computing may find a better use for quantum chemistry than general quantum computing.
The Givens rotation gate are more likely to work correctly than the gates used for quantum computing (at least in the post-select case), and due to the Givens rotations being universal for quantum chemistry~\cite{Arrazola2022universalgivens}, the circuits could succeed more often than circuits from quantum computing gates.
Further investigations are needed for Givens gates that include complex arguments and for controlled Givens gates as well to see if the improvement in success rate is maintained.

We have focussed on results for 2-qubit quantum computing gates, however, finding circuits for gates of other sizes are of interest as well.
Gates that act on a single qubit can be easily found by the tool as these gates can be implemented without any auxiliary wires.
An instance of a larger gate is the Toffoli gate ($CCX$), which acts on three qubits, sending $\ket{\compone\compone\compzero} \to \ket{\compone\compone\compone}$ (and vice versa) and performs the identity on other computational basis states.
There are a few known implementations using linear optics of this and similarly sized gates~\cite{Li2022,Liu23,Alessio24}, but it is unknown if these representations are efficient (in terms of chance of occurring or number of auxiliary/photons used).
Our tool is capable of creating the assertions required to be met by circuits of different sizes, but it would take a long time to return a result/circuit for quantum gates that act on 3 or more qubits except for trivial ones (\eg{} it is \infeasible{} to find a circuit for $CCX$ using no auxiliary wires).

A remark regarding the number of qubits our approach can tackle is in order.
The polynomials derived from the formulas shown in Section~\ref{sec:method:encoding} are very peculiar. First, they have many variables, such as the square of the number of modes. Secondly, their monomials have a low degree. The constraint \texttt{unitary} gives monomials of degree 2, while the constraint \texttt{fockequal} gives monomials of a degree corresponding to the number of photons in the circuit. If for the post-selected CZ of Section~\ref{sec:std-known-postselect} the degree is 2, for heralded gates requiring four photons, the degree increases accordingly. From the perspective of SMT solvers, degree 2 is doable (as shown by our results), but heralded problems quickly reach the limit of what can be done following a naïve approach with existing tools.

Whilst in this paper we have focused on encoding the quantum photonics circuit $\Ulin$ using real variables, it is possible to modify the variables to be complex using the tool and appropriate constraints can be generated.
This involves representing variables as two real variables that represent the real and imaginary part, which essentially doubles the number of variables and constraints needed to specify the properties of $\Ulin$.
This can dramatically increase the time it takes to find a satisfying solution or to prove unsatisfiability.
Whilst it is possible to find unitaries for single qubit operations (\eg the $S$ gate, which does nothing to $\ket{\compzero}$ and sends $\ket{\compone}$ to $\iu\ket{\compone}$), quantum computing operations with two or more qubits currently would take a long time to find a solution or remain out of reach due to the increase in the number of variables.
Work would need to be done by the SMT community to develop a theory and/or an implementation capable of more efficiently solving constraints with complex variables.
Alternatively, developing techniques for handling multi-variable polynomials of limited degree would be helpful for resolving the generated constraints.

Another improvement to the technique is to consider different coincidence basis states can be used, particularly in the auxiliary wires.
Whilst in this paper we have focused on using non-entangled Fock states (\eg $\fock{n_1, \dots, n_m}$), some results have used entangled auxiliary wires as part of the coincidence basis.
For instance, the authors in \cite{Li2022} use the N$00$N state in two auxiliary wires, which is of the form $\frac{1}{\sqrt{2}}(\fock{N,0} + \fock{0,N})$ where $N$ is a positive integer.
Implementing such a feature would increase the variety of photonics circuits to consider.

One final technique to consider is to take advantage of symmetries in the quantum computing gate to reduce the number of variables needed.
The idea is that multiple variables in the photonics gate can be represented by a single variable if certain symmetric properties hold.
For instance, with a $CZ$ gate if an $X$ gate is applied to the qubits, then the $\ket{\compzero\compone}$ and $\ket{\compone\compzero}$ still fundamentally act the same.
This would mean fewer variables and potentially a speed-up in searching for a satisfiable instance.

\section{Conclusion}
\label{sec:conclusion}
In this paper, we have introduced a search technique based on $\delta$-weakening SMT solvers for finding a linear optics circuit that implements a chosen quantum computing gate on a given setup for the circuit.
We showed how any generated approximation from a ($\delta$-weakening) SMT solver that is $\delta$-satisfiable can be turned into an exact solution, which can be used to generate a circuit.
We demonstrated the utility of our technique by demonstrating how our tool can replicate and expand upon results known in the literature of linear optics, and further how it can be used to generate new results.
However, the technique faces a wall in overcoming the heralded setting of linear optics.

This paper highlights important connections between the SMT-based synthesis approach and photonic circuit design for quantum gates.
To overcome the challenge presented by the heralded setting, developments in the area of SMT solving are needed.
In particular, a development of a technique for handling constraints consisting of polynomials of bound degree in the NRA setting would be useful.
Alternatively, development of techniques to solve Complex Arithmetic (CA) theories and implementing them in a tool would be beneficial.

\begin{credits}
\subsubsection{\ackname}
This work has been partially funded by the European Union through the MSCA SE project QCOMICAL, by the French National Research Agency (ANR): projects RECIPROG ANR-21-CE48-0019, PPS ANR-19-CE48-0014, TaQC ANR-22-CE47-0012 and within the framework of ``Plan France 2030'', under the research projects EPIQ ANR-22-PETQ-0007, OQULUS ANR-23-PETQ-0013, HQI-\linebreak[4]{}Acquisition ANR-22-PNCQ-0001 and HQI-R\&D ANR-22-PNCQ-0002.
\end{credits}

\bibliographystyle{splncs04}
\bibliography{ref}

\appendix
\clearpage
\section{Data Tables}
\sisetup{table-auto-round}
\vspace*{-10mm}
\begin{table}[H]
        \centering
        \caption{Table for $CZ$ experiment with a single auxiliary wire (Section~\ref{sec:res:post:onewire}): Number of Photons sent down a single auxiliary wire and the found probability of success (rounded to 4 decimal places) based on an average of successful runs of 5 runs using $\delta = 0.001$ with a 300 second timeout.
        The average is calculated by taking the average of the midpoint of the lower and upper bounds of the success probability for each successful run.
        The average time is calculated based only on the times of the successful runs.
        }
        \label{tab:app:czoneaux}
        \begin{tabular}{|c|S[table-format=2.4]|c|S[table-format=3.4]|}
		\hline
            Number of Photons        & {Average Success Probability} &      Successes       &     {Average Time}    \\
            \hline
            0          &         0.0          &          5           & 0.37128658294677735 \\
            1          & 0.15240442973268786  &          5           &  15.362349796295167 \\
            2          & 0.09624079114404355  &          5           &  111.6834593296051  \\
            3          & 0.07909626948493473  &          5           &  157.01951351165772 \\
            4          & 0.07103398217615617  &          5           &  202.44421916007997 \\
            5          & 0.06640305649841322  &          5           &  241.1577040672302  \\
            6          & 0.06342305001509396  &          5           &  258.3528799533844  \\
            7          &  0.0613548578939378  &          5           &  265.8270726680756  \\
            8          & 0.05985591748840007  &          5           &  281.50336971282957 \\
            9          & 0.05868129082914735  &          5           &  292.2920424938202  \\
            10         & 0.05771803619819313  &          5           &  325.1112949371338  \\
            11         & 0.05699114788261522  &          5           &  331.2661744117737  \\
            12         & 0.05634304204508604  &          5           &  337.8321695804596  \\
            13         & 0.05585752788816514  &          5           &  350.9504230976105  \\
            14         & 0.05539722457399697  &          5           &  409.0192187309265  \\
            15         & 0.05499696028106138  &          5           &  418.18251261711123 \\
            \hline
        \end{tabular}
\end{table}
\vspace*{-10mm}
\begin{table}[H]
        \centering
        \caption{Sample of angles for Givens matrix using two vacuum wires (Section~\ref{sec:res:post:givens}) and their average success probability (rounded down to 4 decimal places) after 5 runs with a $60$ second timeout and $\delta = 0.001$.
        The average time is calculated based only on the times of the successful runs.
        }
        \label{tab:app:givens}
        \begin{tabular}{|c|S[table-format=2.4]|c|S[table-format=2.4]|}
        \hline
        Angle ($a\pi$)     & {Average Success Probability} &      Successes       &     {Average Time}    \\
        \hline
		0/12         &  0.999999941420208   &          5           &  1.7662578105926514 \\
		1/12         &  0.266733756472323   &          4           &  60.53860306739807  \\
		2/12         & 0.24997400763971156  &          3           &  60.50954585075378  \\
		3/12         & 0.17173222800006588  &          4           &  60.53536796569824  \\
		4/12         & 0.13406019450197104  &          4           &  60.52789673805237  \\
		5/12         & 0.11641232595068461  &          4           &  60.51867203712463  \\
		6/12         & 0.11125014702452933  &          5           &  60.55566296577454  \\
		7/12         & 0.11641148493142597  &          3           &  60.51067667007446  \\
		8/12         & 0.13403508571599682  &          2           &  60.49877190589905  \\
		9/12         &  0.1716830187631701  &          4           &  60.50402207374573  \\
		10/12         & 0.25000826058259884  &          4           &  60.51524701118469  \\
		11/12         &  0.2625539211884096  &          3           &   60.5392418384552  \\
		12/12         &  0.9999999999999998  &          5           &  1.2743714332580567 \\
		\hline
        \end{tabular}
\end{table}
\clearpage


\end{document}